\documentclass[useAMS]{mn2e} 

\usepackage{txfonts} 
\usepackage{multirow}
\usepackage{fmtcount}
\usepackage{graphicx}

\newcommand{\epeiso}{$E_{\rm peak}-E_{\rm iso}$}
\newcommand{\epliso}{$E_{\rm peak}-L_{\rm iso}$}
\newcommand{\ep}{$E_{\rm peak}$}
\newcommand{\epo}{$E^{\rm obs}_{\rm peak}$}
\newcommand{\eiso}{$E_{\rm iso}$}
\newcommand{\liso}{$L_{\rm iso}$}

\title[Test for spectral-energy correlations]
{A complete sample of bright \emph{Swift} Long Gamma--Ray Bursts: testing the spectral--energy correlations}

\author[L. Nava et al.]
{L.~Nava,$^{1}$\thanks{E-mail: lara.nava@sissa.it} 
R.~Salvaterra,$^2$
G.~Ghirlanda,$^3$
G.~Ghisellini,$^3$
S.~Campana,$^3$
S.~Covino,$^3$
\newauthor
G.~Cusumano,$^4$
P.~D'Avanzo,$^3$
V.~D'Elia,$^{5,6}$
D.~Fugazza,$^3$
A.~Melandri,$^3$
B.~Sbarufatti,$^3$
\newauthor
S.~D.~Vergani$^3$
and G.~Tagliaferri$^3$\\
$^1$SISSA - Via Bonomea 265, I-34136 Trieste - Italy\\
$^2$INAF - IASF Milano, via E. Bassini 15, I-20133 Milano, Italy \\
$^3$INAF - Osservatorio Astronomico di Brera, via E. Bianchi 46, I-23807 Merate, Italy\\ 
$^4$INAF - IASF Palermo, Via Ugo La Malfa 153, I-90146 Palermo, Italy\\
$^5$ASI-Science Data Centre, Via Galileo Galilei, I-00044 Frascati, Italy\\
$^6$INAF - Osservatorio Astronomico di Roma, via di Frascati 33, 00040 Monte Porzio Catone, Italy
}
\begin{document}

\date{}

\pagerange{\pageref{firstpage}--\pageref{lastpage}} \pubyear{2002}

\maketitle

\label{firstpage}

\begin{abstract}
We use a  nearly complete sample of Gamma Ray Bursts (GRBs) detected by the {\it Swift} satellite
to study the correlations between the spectral peak energy \ep\ of the prompt emission, 
the isotropic energetics \eiso\ and the isotropic luminosity \liso. 
This GRB sample is characterized by a high level of completeness in redshift (90\%).
This allows us to probe in an unbiased way the issue related to the physical origin of 
these correlations against selection effects. 
We find that one burst, GRB 061021, is an outlier to the \epeiso\ correlation.
Despite this case,
we find strong \epeiso\ and \epliso\ correlations for the bursts of the complete sample. 
Their slopes, normalizations and dispersions are consistent with those found with the whole 
sample of bursts with measured redshift and \ep. 
This means that the biases present in the total sample commonly used to study these correlations 
do not affect their properties.  Finally, we also find no evolution with redshift of the \epeiso\ and \epliso\ correlations.
\end{abstract}

\begin{keywords}
gamma-ray: burst --
\end{keywords}

\section{Introduction}

One of the most debated issues in Gamma Ray Bursts (GRBs) concerns the existence of correlations 
among the spectral parameters of the prompt emission and its energetic and luminosity. 
Three robust correlations have been identified. 
Each of them involves the rest frame peak energy \ep\ of the $\nu F_{\nu}$ prompt spectrum. This quantity strongly 
correlates with (i) the isotropic energy \eiso\ (Amati relation, Amati et al. 2002), (ii) the isotropic peak luminosity \liso\ 
(Yonetoku relation, Yonetoku et al. 2004) and (iii) the collimation--corrected energy $E_\gamma$ (Ghirlanda 
relation, Ghirlanda et al. 2004b; see also Liang \& Zhang 2005). 
These correlations are valid for long GRBs. There are indications that short bursts (i.e. bursts with observed duration 
$<$2 s) obey the very same \epliso\ relation defined by long events but they are inconsistent with the \epeiso\ 
correlation (Ghirlanda et al. 2009; Amati 2010).

The spectral--energy correlations have relevant implications both for the theoretical understanding of the burst 
physics and for the application of GRBs as cosmological tools (Ghirlanda et al. 2004a; Ghirlanda et al. 2006).  
While there is no consensus yet on the interpretation of these correlations (e.g. Eichler \& Levinson 
2004; Levinson \& Eichler 2005; Rees \& M\'esz\'aros 2005; Thompson, M\'esz\'aros \& Rees 2007), 
an interesting progress has been recently made by 
studying the comoving properties of GRBs (Ghirlanda et al. 2012; Panaitescu et al. 2009).

The \epeiso\ and \epliso\ correlations have been discovered with a very small sample of 12--16 GRBs detected by  {\it Beppo}SAX and/or BATSE 
(Amati et al. 2002; Yonetoku et al. 2004). They have been subsequently confirmed by adding bursts detected 
by other satellites (e.g. {\it HETE II}, {\it Konus}/WIND, {\it INTEGRAL}, {\it Swift} and {\it Fermi}). 
At present the sample of bursts defining 
the \epeiso\ and \epliso\ correlations comprises more than 130 events, with only two well known (very peculiar) 
outliers (but see Ghisellini et al. 2006). 

Despite the considerably large number of bursts consistent with these correlations, their physical origin 
is debated. Some authors claim that they are the result of instrumental selection effects  
(Nakar \& Piran 2005; Band \& Preece 2005; Butler et al. 2007; Shahmoradi \& Nemiroff 2011). 
Other studies (Ghirlanda et al. 2008; Nava et al. 2008) 
quantified the possible instrumental selection biases finding that, even if they do affect the sample, they cannot be 
responsible for the existence of the spectral--energy correlations. 
Ghirlanda et al. (2010; 2011a; 2011b) added an important piece of information to this debate. They analysed the
time--resolved spectra of both long and short bursts detected by the {\it Fermi} satellite. They found that the 
time--resolved \ep\ correlates with the luminosity during the temporal evolution of the burst (and  
for bursts without $z$ they derive a correlation between the time--resolved \epo\ and the flux). 
This is a feature common to short and long GRBs. The time--resolved correlations found within individual
GRBs are similar to each other and they are similar to the \epliso\ relation defined by the time--integrated 
spectral properties of different bursts (i.e. the Yonetoku relation). 
This is a strong argument in favour of the physical origin of the correlations because it is very unlikely that within a 
single burst the instrumental selection effects (e.g. the detector threshold) or possible effects related to the dependence 
of the peak energy, energetic and luminosity on the redshift can play any role.

GRBs added to the \epeiso\ and \epliso\ correlations need to have their redshifts measured. This raised the suspect that 
these correlations might be biased by the high level of incompleteness in redshift of the samples of GRBs so far used. 
In this paper we study, for the first time, the  \epeiso\ and  \epliso\ correlations with a sample of GRBs with a high 
level of completeness (90\%). 
The sample is presented by Salvaterra et al. (2012) and it comprises 58 events detected by  {\it Swift} with a 1--s peak 
photon flux 
$P\geq2.6$ ph s$^{-1}$cm$^{-2}$ (integrated in the 15--150 keV energy range).  
Fifty--two out of 58 GRBs have measured redshift $z$, while in 3 cases an upper limit on $z$ can be set from the photometry of the afterglow or of the host galaxy.
The completeness level then increases up to $\sim$95\% by considering the redshift constraints imposed by the detection of the afterglow or host galaxy 
in some optical filters.

The complete sample used in this work is particularly appropriate to study the possible evolution of the 
\epeiso\ and \epliso\ correlations with redshift. 
Li (2007), considered a sample of 48 GRBs (from Amati 2006, 2007) and investigated if the \epeiso\ correlation 
evolves with redshift, finding that it does, becoming steeper at higher redshifts. Ghirlanda et al. (2008) repeated the 
same test with a larger sample of 76 bursts and found no evidence of evolution with redshift. 
These results, however, were obtained by considering highly incomplete GRB samples. 

Our goals are (i) to study how \ep\ is correlated to \eiso\ and \liso\ within the complete sample of bursts and (ii) to 
study the possible evolution of these correlations with redshift through this unbiased sample. 
In $\S$2 we present the spectral properties of the GRBs composing the complete sample. 
We study the \epeiso\ and \epliso\ correlations defined by this sample in $\S3$ and compare their properties with 
those defined by the (incomplete) larger sample of GRBs with measured $z$ and \epo.
In $\S4$ we discuss and summarise our conclusions. 
Throughout the paper we assume a standard cosmology with $h=\Omega_{\Lambda}=0.7$ and $\Omega_{m}=0.3$.

\section{The sample}

In order to study a relatively unbiased redshift distribution of GRBs, Jakobsson et al. (2006) carefully selected a 
sample containing bursts which have `observing conditions' favourable for redshift determination (see their web 
page{\footnote{http://www.rauvis.hi.is/$\sim$pja/GRBsample.html}} for an updated compilation of the sample and 
an explanation of the selection criteria). 
This relatively `clean' sample has a redshift recovery rate of roughly 50\%. 
In order to reach a higher level of completeness, Salvaterra et al. (2012) considered this sample and also required 
that the photon peak flux, measured by {\it Swift}/BAT in the 15--150 keV energy band, is $P\geq2.6$ ph 
s$^{-1}$cm$^{-2}$.

Up to May 2011, 58 GRBs match this selection criteria and are listed in Table \ref{tab1}. 
Fifty-two of them have measured redshift so that our completeness level is 90\%.
Of these 52, all but two (namely GRB 070521 and GRB 080602) 
have spectroscopic confirmed redshift either from absorption lines over-imposed on the GRB optical afterglow 
or from emission lines of the GRB host galaxy. Moreover, for 3 of the 6 bursts lacking measured $z$ the 
afterglow or the host galaxy have been detected in at least one optical filter, so that $\sim$95\% 
of the bursts in this sample have a constrained redshift. While this sample 
represents only $\sim$10\% of the full Swift sample, it contains more than 30\% of long 
GRBs with known redshift. 

The bursts belonging to this sample (updated to May 2011) are listed in Table \ref{tab1}. 
We will refer to this sample as the `complete' one.

For all bursts in this sample,
we collected from the literature the prompt emission spectral parameters and
computed the rest frame peak energy \ep=\epo$(1+z)$, the isotropic energy \eiso\ and the isotropic peak 
luminosity \liso. 
The two latter quantities have been calculated in the rest frame energy range 1 keV--10 MeV. 
The narrow sensitivity range of the BAT instrument allows to determine \epo\ only when it falls approximately  
between $\sim15$ keV and $\sim$150 keV. 
This condition is necessary but not sufficient: as demonstrated 
in the spectral simulation studies by Sakamoto et al. (2009), even when \epo\ lies in the BAT energy range, the value of \epo\ can be determined only if the spectrum is characterized by a high signal-to-noise. Similar conclusions were reached through numerical simulations by Ghirlanda et al. (2008) and Nava et al. (2008).
Almost all GRBs of the complete sample, however, have been detected also 
by other instruments with wider sensitivity ranges (mainly {\it Konus}/WIND and, for bursts after August 2008, {\it Fermi}). 
This allowed to estimate \epo\ for a large fraction of events.
We, therefore, obtained a sample of 46 bursts (out of 58) with  \ep, \eiso, \liso\ and redshift $z$, which are 
the quantities required to investigate the \epeiso\ and \epliso\ correlations. 
For 6 GRBs there is a good knowledge of the prompt spectral shape (i.e. \ep\ is well constrained), but the redshift is 
unknown (in 3 cases it is possible to derive an upper limit on $z$ from the photometry of the afterglow or of the host galaxy). 
For the remaining 6 bursts the determination of \epo\ is not possible: in 3 cases the spectrum is well described by a 
single power--law, while in the other 3 cases it is possible to 
set a lower limit to the observed value of \epo. 
Redshifts and spectral properties of the 58 GRBs of the complete sample are reported in Table \ref{tab1}.
Here we also report the spectral parameters of the prompt emission spectrum (with their uncertainties) and the fluence and peak flux 
(with the energy ranges over which they are computed). These are the informations used to compute the rest frame isotropic energy and luminosities of the
bursts in our samples. 
We also report for each burst the instrument used to derive the spectral parameters listed in the table.
To summarize, our sample is composed as follows:
\begin{itemize}
\item 46 GRBs have measured $z$ and \ep\ (and consequently \eiso\ and \liso);
\item 3 GRBs have measured \ep\ but only an upper limit on $z$;
\item 3 GRBs have measured \ep\ but no redshift estimate;
\item 3 GRBs have measured $z$ but only a lower limit on \epo;
\item 3 GRBs have measured $z$ but no \epo\ estimate;
\end{itemize}

It is interesting to compare the \epeiso\ and \epliso\ correlations defined by the complete sample with the same 
correlations defined with a larger sample of GRBs, comprising all bursts with measured redshifts and \epo\ 
detected by different satellites. 
We updated also this sample (called in the following the `total' sample) to May 2011. 
The total sample contains 136 bursts, for one of which the peak luminosity cannot be derived, due to the lack of 
measured peak flux. 
For GRBs detected by the Gamma ray Burst Monitor (GBM -- Meegan et al. 2009) onboard {\it Fermi} after March 
2010 we retrieved the public data from the online catalog and we analysed their time--integrated spectra following 
the standard procedure described in Nava et al. (2011a). 
This sample is based on the one presented in Ghirlanda et al. (2012) to which we added four
bursts (GRB 060306, GRB 
061021, GRB 081007, GRB 081221 belonging also to the complete sample - see Table \ref{tab1}). 

We can thus compare the spectral--energy correlations defined by the complete and by the total samples. 
To point out possible differences 
we also consider the sample obtained from the total one by excluding the bursts of the complete sample (we will 
refer to it as the `complementary' sample).

\section{Results}
\begin{figure*}
\includegraphics[scale=0.57]{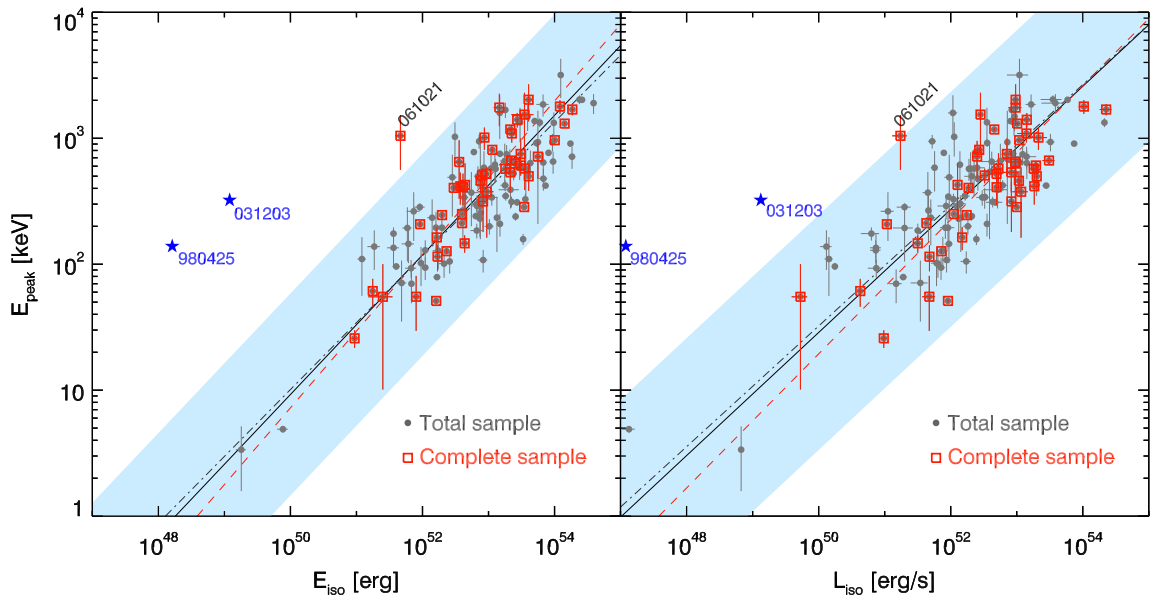}
 \caption{Amati (left panel) and Yonetoku (right panel) correlations. Grey filled circles refer to the total sample.
  Their power--law fit is shown as a solid dark line. 
 The shaded region represents the 3$\sigma$ scatter of the distribution of points around this best fit line. 
 Bursts belonging to the complete sample (and with secure redshift and \ep\ estimate -- 46
  events) are marked as 
 empty red squares and the red dashed line represents their best fit model. The dot--dashed line is the fit to the 
 complementary sample.
 The position of the two historical outliers, GRB 980425 and GRB 031203, is also shown (blue stars), together with the
 burst
  of our complete sample, GRB 061021, that is
   above the $3 \sigma$ scatter limit.
 }
 \label{fig1}
\end{figure*}
\begin{figure*}
\includegraphics[scale=0.57]{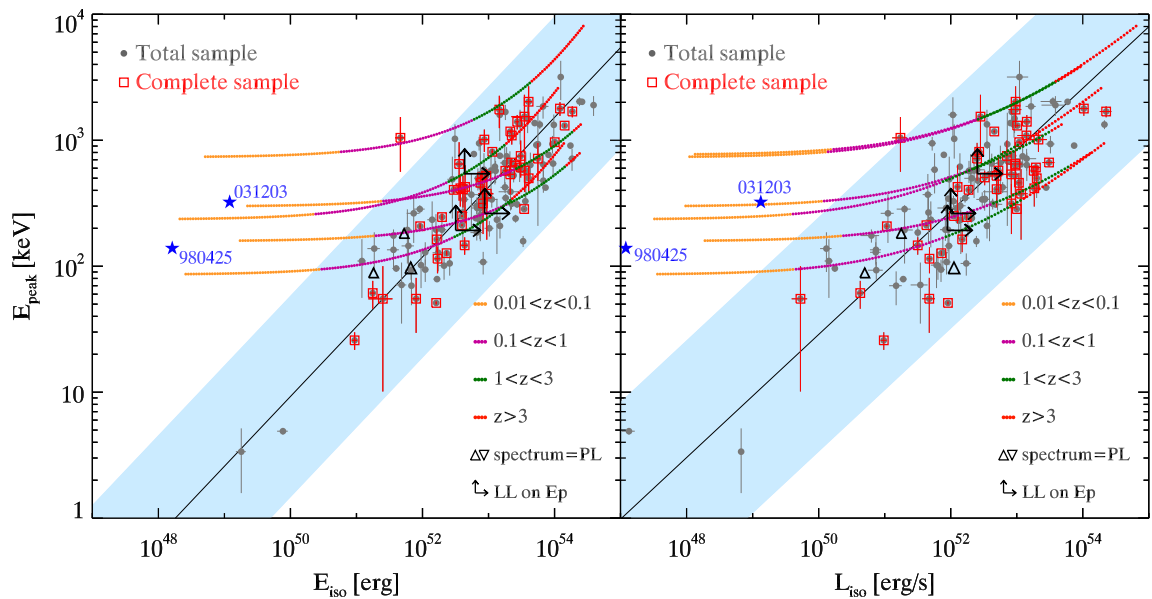}
 \caption{The consistency with the Amati (left panel) and Yonetoku (right panel) correlations of bursts with unknown 
 redshift or \epo\ is shown. 
 For bursts without redshift estimate (or with an upper limit on the redshift) the test is performed by varying $z$ from 
 0.01 to 10 (or to the upper limit). 
 The resulting \ep, \eiso\ and \liso\ are shown as continuous lines. 
 Different colours refers to different ranges of redshift (see legend). 
 Arrows mark those bursts for which $z$ is known but there is only a lower limit on the \epo\ value (which transforms 
 into a lower limit on \eiso\ and \liso).
 Triangles represent those events whose spectrum is well described by a single power--law model. 
 In these cases we can set a lower limit (upwards triangles) on \ep, \eiso\ and \liso\ by 
 considering the photon index of the power--law fit and the energy range of sensitivity of the instrument. The total 
 sample (and its 3$\sigma$ scatter -- shaded region) and the complete sample are also shown for reference. 
 The position of the two historical outliers, GRB 980425 and GRB 031203, is also shown (blue stars).}
 \label{fig2}
\end{figure*}
\begin{figure*}
\includegraphics[scale=0.57]{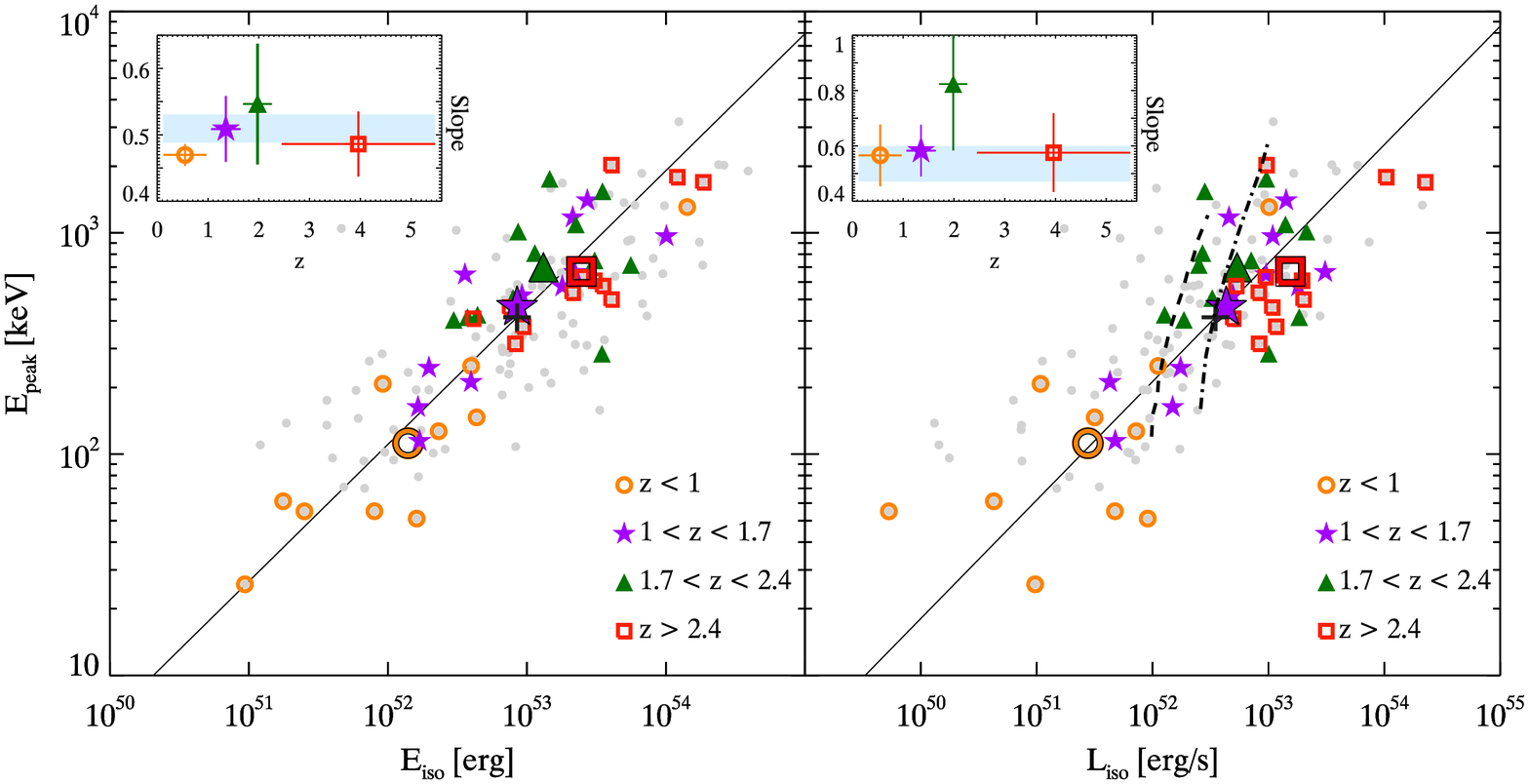} 
 \caption{Evolution with redshift for both the \epeiso\ (left panel) and the \epliso\ (right panel) correlation. 
 GRBs from the complete sample have been divided into 4 sub--samples, accordingly to their redshift (see legend).
 The inset shows the slope of the correlation for each sub--sample.
 The shaded blue region is the two sided 1$\sigma$ interval of the slope found by considering the whole complete 
 sample of 46 GRBs and by excluding for consistency GRB 061021. Big symbols mark the centre of mass of data points for each subsample and for the whole sample (dark cross). Grey filled circles show data points from the total sample. The dashed black 
 curve marks the lowest possible luminosity of the flux limited sample estimated for GRBs with $z=2$. The dot--dashed 
 curve represents the same for bursts with $z=3$.}
 \label{fig3}
\end{figure*}

We first test the \epeiso\ and the \epliso\ correlations with the sample of  46
bursts (included in the complete sample) 
with firm estimates of the redshift and of the spectral properties. 
The results are shown in Fig. \ref{fig1}. 
Data points from the total sample are plotted for comparison as grey dots. 
Red squares show the bursts of the complete sample. We estimate the Spearman's rank correlation coefficient 
and the associated chance probability for the complete, total and complementary sample and report these 
numbers in Table \ref{tab2}. We model the distribution of the data points in the logarithmic plane of Fig. \ref{fig1} 
with a linear function. We applied the ordinary least squares bisector method (Isobe et al. 1990) for fitting the data. 
This choice is motivated by the fact that a priori there is no reason to assume either \ep\ or \eiso\ (or \liso) as the 
independent variable, and also by the large scatter of the data points.  
The errors on the slope and normalization are determined by fitting in the barycenter of data points, where they are uncorrelated. The results are given in Table \ref{tab2}. 
In Fig. \ref{fig1} the best fit lines are shown for each sample as follows: a solid black line for the total sample, a 
dashed red line for the complete sample and a dot--dashed black line for the complementary sample. 
The shaded region marks the 3$\sigma$ of the scatter distribution of the data points around the best fit line for the 
total sample. The scatter estimated for the other two samples is also listed in Table \ref{tab2}.
In Fig. \ref{fig1} we also report the position of GRB 980425 and GRB 031203 (blue filled stars), which are well known outliers 
of both the \epeiso\ and the \epliso\ correlation. 
These two bursts have peculiar properties and they are generally considered as outliers to the confirmed relations.

The behaviour of GRB 061021 is very peculiar, since it lies at more than the 3$\sigma$ from the Amati and Yonetoku correlations 
(considering the scatter defined by the total sample -- shaded regions in Fig. \ref{fig1}). 

Twelve events belonging to the complete sample cannot be directly used to test the correlations, since they have 
some unknown property: the redshift or the spectral peak energy (see Sec. 2). 
However, we can still test their consistency with the correlations (results are shown in Fig. \ref{fig2}). 
To check the consistency 
different methods are adopted for the following different cases:

\begin{itemize}
\item 
lower limit on \epo\ and secure estimate of $z$ (3 cases): we derive the lower limit on the rest frame peak 
energy and on \eiso\ and \liso\ (right and up arrows in Fig. \ref{fig2}); 

\item 
no estimate of \epo\ and secure estimate of $z$ (3 cases): we consider the energy range of sensitivity of the 
instrument and the photon index of the power--law fit. 
Photon indices greater (lower) than --2 [in the notation $N(E) \propto E^{\Gamma}$] suggest peak 
energies greater (lower) than the upper (bottom) energy edge of the instrument. 
An upper or lower limit to \ep\ (and consequently on \eiso\ and \liso) can be set 
(empty triangles in Fig. \ref{fig2}); 

\item 
spectral properties well constrained but upper limit on $z$ (3 cases): \ep, \eiso\ and \liso\ are estimated for 
a variable redshift ranging from 0.01 to its upper limit (coloured continuous curves in Fig. \ref{fig2}); 

\item 
spectral properties well constrained but no redshift estimate (3 cases): \ep, \eiso\ and \liso\ are estimated for 
values of the redshift ranging from 0.01 to 10 (coloured continuous curves in Fig. \ref{fig2}).
\end{itemize}

\subsection{Evolution with redshift}

The possibility that evolutionary effects play a role in the spectral--energy correlations still represents an open issue. 
To investigate if the slopes of the \epeiso\ and \epliso\ correlations have a dependence on the redshift we repeat the 
same method used by Li (2007) and Ghirlanda et al. (2008) and apply it to our complete sample. 
We exclude from the fit the burst which lies at more than 3$\sigma$
of the Amati and Yonetoku relations. We define 4 
sub--samples, with almost the same number of bursts, divided as follows: 
(i) $z<1$ (10 events, yellow empty circles in Fig. \ref{fig3}), 
(ii) $1<z<1.7$ (11 events, filled purple stars), 
(iii) $1.7<z<2.4$ (12 events, filled green triangles) and 
(iv) $z>2.4$ (12 events, empty red squares). 
Bursts belonging to the total sample  are not used for this test,
since we want to take advantage of the completeness of our sample. 
The best fit slope of each sub--sample is reported in the inset of Fig. \ref{fig3}, 
both for the Amati (left panel) and Yonetoku (right panel) correlation. 
The light blue horizontal stripe, shown in the inset, represents the two--sided 1$\sigma$ interval of 
the slope derived by considering the whole complete sample and by excluding (for consistency) the two outliers. 
There is no evidence of a systematic evolution of the slope  with $z$. 
The slopes defined by individual bins of redshift are all consistent, within 1$\sigma$, with the slope 
defined by the entire complete sample. This result does not change by choosing a different number of bins.

Bursts with the highest redshifts populate only the high--energy (and high--luminosity) region of 
the Amati and Yonetoku correlations. This is due to the fact that we selected relatively bright bursts,
by requiring a minimum peak flux. The effect of this selection can be investigated as follows. 
We consider a series of spectra characterised by typical values $\alpha\sim -1$ and $\beta\sim-2.3$. 
We impose that the photon peak flux in the 15--150 keV energy range is equal to our limiting flux 
($P=2.6$ ph s$^{-1}$cm$^{-2}$). For different values of \epo\ we derive the corresponding 
limiting luminosity by considering two cases: (i) $z=2$ (dashed curve in 
Fig. \ref{fig3}) and (ii) $z=3$ (dot--dashed curve). 
We consider these redshifts since they are representative of the two sub--samples at higher redshifts, 
which populate only the high--energy (and high--luminosity) part of the planes. From Fig. \ref{fig3} 
it is clear that this effect is due to the use of a flux limited sample. 

We may wonder if these limiting luminosities derived at different redshifts may affect our study on the redshift evolution. We identify two possible effects: at high redshift the range of available luminosities is smaller, i.e. the reduced leverage in luminosity might harden the estimate of the correlation slope and enlarge the associated uncertainty. Moreover, since the introduced cutoff is nearly vertical in the \ep--\liso\ plane, we expect a possible steepening of the slopes toward high redshift. This effect has been investigated also by Li (2007). By performing simulations, he concludes that a spurious evolution of the slope with the redshift can be introduced, with high-redshift bins defining steeper relations. We do not find signs of this effect in our data (see the inset in Fig. \ref{fig3}), indicating that, even if at high redshift we are missing low-luminosity events, we are still able to recover the slope of the correlations which are consistent with those derived with low--z bins.


\begin{table*}
 \centering
 \begin{minipage}{176mm}
  \caption{List of bursts that define the complete sample. Redshifts (from Salvaterra et al. 2012),  spectral parameters, rest frame peak energies \ep, isotropic energies \eiso\ and luminosities \liso\ are provided. \eiso\ and \liso\ are estimated in the 1-10$^4$ keV energy range. All quoted errors are at the 90\% confidence level. Column 8 reports the name of the mission from which spectral properties have been derived: S=Swift, K=Konus/Wind, F=Fermi, Su=Suzaku, H=Hete. References in the last columns are for the spectral properties: [1] Perri et al. 2005, [2] Sakamoto et al. 2008, [3] Golenetskii et al. 2005a, [4] Sakamoto et al. 2006, [5] Blustin et al. 2005, [6] Butler et al. 2007, [7] Sakamoto et al. 2011b, [8] Crew et al. 2005, [9] Golenetskii et al. 2005b, [10] Cabrera et al. 2007, [11] Sakamoto et al. 2006b, [12] Barthelmy et al. 2006, [13] Amati et al. 2007b, [14] Golenetskii et al. 2006a, [15] Sakamoto et al. 2011, [16] Golenetskii et al. 2006b, [17] Golenetskii et al. 2006c, [18] Golenetskii et al. 2007a, [19] Golenetskii et al. 2007b, [20] Krimm et al. 2007, [21] Golenetskii et al. 2007c, [22] Racusin et al. 2008, [23] Golenetskii et al. 2008a, [24] Golenetskii et al. 2008b, [25] Golenetskii et al. 2008c, [26] Golenetskii et al. 2008d, [27] Golenetskii et al. 2008e, [28] Golenetskii et al. 2008f, [29] Starling et al. 2009, [30] Nava et al. 2011a, [31] Bissaldi et al. 2008, [32] Golenetskii et al. 2008g, [33] Ukwatta et al. 2008, [34] Golenetskii et al. 2009a, [35] Golenetskii et al. 2009b, [36] Golenetskii et al. 2009c, [37] Pal'shin et al. 2009, [38] Noda et al. 2009, [39] Golenetskii et al. 2009d, [40] Foley et al. 2010, [41] Golenetskii et al. 2010, [42] this work, [43] Golenetskii et al. 2011a, [44] Golenetskii et al. 2011b.}
\addtocounter{footnote}{0}
\footnotetext[\value{footnote}]{In these cases the peak flux is in units of ph/cm$^2$/s}
\addtocounter{footnote}{1}
\footnotetext[\value{footnote}]{For GRBs without measured $z$ the listed peak energy refers to the observed one \epo.}
\addtocounter{footnote}{1}
\footnotetext[\value{footnote}]{The spectrum is well described by a power--law model and \epo\ is uncostrained.}
\addtocounter{footnote}{-2} 
\begin{center}
\tabcolsep 4.5pt
\begin{tabular}{@{}lcccccccccccl@{}}
  \hline
GRB & $z$ &  $\alpha$ [$\beta$] & Fluence & Range & Peak Flux & Range & Mission & \ep\ & \eiso\ & \liso\  &Ref. \\
         &        &                             & 10$^{-6}$erg/cm$^{2}$ &  keV & erg/cm$^{2}$/s&keV & & keV&10$^{52}$erg&10$^{51}$erg/s& \\
  \hline
050318  &   1.44  & -1.34$\pm$0.32  &   2.1$\pm$0.21 &  15-350   &  (2.2$\pm$0.17)$\times10^{-7}$  & 15-150 & S & 115$\pm$27   &  1.69$\pm$0.17  &   4.76$\pm$0.37  &1, 2 \\
050401  &   2.90  & -1.0    [-2.45]  &   19.3$\pm$0.4 &  20-2000  &  (2.45$\pm$0.12)$\times10^{-6}$ & 20-2000 & K & 499$\pm$117   &  40.6$\pm$0.84  &   201$\pm$9.85  &3 \\
050416A &   0.653 & -1.0   [-3.4]  &   0.35$\pm$0.03 &  15-150   &  5.0$\pm$0.5$^{\alph{footnote}}$     & 15-150    & S&  26$\pm$4    &  0.094$\pm$0.008  &   0.97$\pm$0.09 &4  \\
050525A &   0.606 & -0.99$\pm$0.11   &   20.1$\pm$0.5 &  15-350	&  47.7$\pm$1.2$^{\alph{footnote}}$    & 15-350	& S &127$\pm$6    &  2.32$\pm$0.06 &   7.23$\pm$0.18  &5 \\
050802  &   1.71  & -1.6$\pm$0.1    &   2.7$\pm$0.4  &  15-350	&  (22.1$\pm$3.53)$\times10^{-8}$ & 15-150	& S&   $>$192     &  $>$3.19	  &   $>$8.90     &6, 7\\
050922C &   2.198 & -0.83$\pm$0.24   &   3.14$\pm$0.31 &  2-400	&  (4.5$\pm$0.7)$\times10^{-6}$  & 20-2000	& H/K&416$\pm$118  &  3.74$\pm$0.37  &   184$\pm$28.7 &8,9 \\
060206   &   4.048 & -1.12$\pm$0.30   &   0.84$\pm$0.04 &  15-150	&  (2.02$\pm$0.13)$\times10^{-7}$ & 15-150&S& 409$\pm$116  &  4.10$\pm$0.21  &   49.6$\pm$3.24  &7 \\
060210  &   3.91  & -1.12$\pm$0.26 &     6.92$\pm$0.37 &  15-150	&  2.8$\pm$0.3$^{\alph{footnote}}$     & 15-150	& S&574$\pm$187  &  35.3$\pm$1.9  &  52.8$\pm$5.66  & 10,11\\
060306  &  3.5 & -1.2 $\pm$0.5  &      2.5$\pm$0.34 &  15-350	&  (47.1$\pm$2.78)$\times10^{-8}$ & 15-150&S&  315$\pm135$       &  8.26$\pm$1.12 	  &    83.0$\pm$4.9  &6, 7	\\
060614  &   0.125 &         &      21.7$\pm$0.4 &  15-150	&  11.6$\pm$0.7$^{\alph{footnote}}$\addtocounter{footnote}{1}     & 15-150	& S& 55$\pm$45   &  0.25$\pm$0.10 &   0.053$\pm$0.014 &12,13 \\
060814  &   1.92  & -1.43$\pm$0.16 &     26.9$\pm$2.6 &  20-1000  &  (2.13$\pm$0.35)$\times10^{-6}$ & 20-1000&K& 750$\pm$245  &  30.7$\pm$2.97)  &   70.9$\pm$11.7 &14  \\
060904A &	  & -1.22$\pm$0.05 &   8.49$\pm$0.50 &  15-1000& (1.3$\pm$0.3)$\times10^{-6}$  & 20-10000 &K/Su/S& 235$\pm$25$^{\alph{footnote}}$\addtocounter{footnote}{1}  &  &  &15 \\
060908  &   1.88  & -0.93$\pm$0.25 &     2.81$\pm$0.11 &  15-150	&  (2.81$\pm$0.23)$\times10^{-7}$ & 15-150	&S & 426$\pm$207  & 4.41$\pm$0.18	&   12.7$\pm$1.04 &7 \\
060912A$^{\alph{footnote}}$  &   0.94  & -1.85$\pm$0.08   &   2.87$\pm$0.42 &  15-1000 &	 (2.5$\pm$0.9)$\times10^{-6}$	&  20-10000&K/Su/S  & & & &15 \\
060927  &   5.47  & -0.81$\pm$0.36    &  1.12$\pm$0.07 &  15-150   &  (2.47$\pm$0.17)$\times10^{-7}$ & 15-150 &S & 459$\pm$90   &  7.56$\pm$0.46 &    108$\pm$7.6  &7  \\
061007  &   1.261 & -0.75$\pm$0.02 [-2.79$\pm$0.09]&  193$\pm$2.62 &  15-1000  &  (1.2$\pm$0.1)$\times10^{-5}$  & 20-10000 &K/Su/S & 965$\pm$27&101$\pm$1.4 &   109$\pm$9.1 &15 \\
061021  &   0.346 & -1.22$\pm$0.13  	&  13.4$\pm$2.3  &  20-2000  &  (3.72$\pm$0.93)$\times10^{-6}$ & 20-2000&K& 1046$\pm$485 &  0.46$\pm$0.08  &    1.73$\pm$0.43 &16   \\
061121  &   1.314 & -1.32$\pm$0.05  	&  56.7$\pm$3.9 &  20-5000  &  (1.28$\pm$0.17)$\times10^{-5}$ & 20-5000&K& 1402$\pm$185 &  27.2$\pm$1.87  &    142$\pm$18.9  &17   \\
061222A &   2.09  & -1.00$\pm$0.05 [-2.32$\pm$0.38]&  16.2$\pm$0.68  & 15-1000  &  (4.8$\pm$1.3)$\times10^{-6} $ & 20-10000&K/Su/S & 1091$\pm$167 &  22.5$\pm$0.94 & 140$\pm$38 &15   \\
070306  &   1.50  & -1.67$\pm$0.1      &  9.0$\pm$0.63 &  15-350   &  (30.4$\pm$1.64)$\times10^{-8}$ & 15-150  &S & $>$263   &  $>$8.74 	 &    $>$9.99   &6,7 \\
070328  &   $<$4 & -1.11$\pm$0.04 [-2.33$\pm$0.24]&  32.6$\pm$1.24 &  15-1000 &  (5.9$\pm$1.2)$\times10^{-6} $ & 20-10000&K/Su/S &  $<3835$ &  $<133$ &  $<690$&15    \\
070521  &   1.35  & -0.93$\pm$0.12 	&  18.1$\pm$1.4 &  20-1000  &  (4.12$\pm$0.91)$\times10^{-6}$ & 20-1000&K& 522$\pm$56   &  9.22$\pm$0.71  &   49.3$\pm$10.9 &18 \\
071020  &   2.145 & -0.65$\pm$0.29 &   7.71$\pm$1.36&  20-2000  &  (6.04$\pm$2.08)$\times10^{-6}$ & 20-2000 &K	& 1013$\pm$204 &  8.65$\pm$1.53 &    213$\pm$73 &19    \\
071112C$^{\alph{footnote}}$\addtocounter{footnote}{-2}  & 0.82  &  -1.09$\pm$0.07 &   3.0$\pm$0.4 &   15-150&	 8$\pm$1$^{\alph{footnote}}$\addtocounter{footnote}{2} &  15-150 &S &	 & 	&   &20     \\
071117  &   1.331 & -1.53$\pm$0.15   &  5.84$\pm$0.85 &  20-1000  &  (6.66$\pm$1.83)$\times10^{-6}$ & 20-1000 &K &  648$\pm$317 &  3.59$\pm$0.52 &    95.3$\pm$26.2  &21 \\
080319B &   0.937 & -0.86$\pm$0.01 [-3.59$\pm$0.45]&  613$\pm$13 &  20-7000  &  (2.26$\pm$0.21)$\times10^{-5}$ & 20-7000 &K & 1307$\pm$43  &  142$\pm$3  &    102$\pm$9.4  &22   \\
080319C &   1.95  & -1.20$\pm$0.10 &  15.0$\pm$2.7 &  20-4000  &  (3.35$\pm$0.74)$\times10^{-6}$ & 20-4000 &K & 1752$\pm$504 &  14.6$\pm$2.6  &    96.1$\pm$21.2 &23     \\
080413B &   1.10  & -1.23$\pm$0.25 &   3.25$\pm$0.13 &  15-150   &  (14.0$\pm$0.58)$\times10^{-7}$ & 15-150 &S & 163$\pm$34   &  1.65$\pm$0.06 &    14.9$\pm$0.62  &7	\\
080430$^{\alph{footnote}}$\addtocounter{footnote}{-1}  &   0.77 &-1.73$\pm$0.08  & 1.17$\pm$0.05 &  15-150&	 (18.2$\pm$1.31)$\times10^{-8}$ & 15-150  &S&    & 	&  &7     \\
080602  &$\sim$1.4& -0.96$\pm$0.63   &  7.92$\pm$1.46 &  20-1000  &  (1.92$\pm$0.58)$\times10^{-6}$ & 20-1000 &K  & $>$542	  &   $>$4.34     & $>$25.3     &24 \\
080603B & 2.69    & -1.23$\pm$0.64   &  4.50$\pm$1.17 &  20-1000  &  (1.51$\pm$0.39)$\times10^{-6}$ & 20-1000 &K & 376$\pm$214  &  9.41$\pm$2.45  & 116$\pm$30     &25 \\
080605  & 1.64    & -1.03$\pm$0.07 &     30.2$\pm$1.2 &  20-2000  &  (1.60$\pm$0.33)$\times10^{-5}$ & 20-2000 &K &  665$\pm$48  &  22.1$\pm$0.88 &    308$\pm$64&26 \\
080607  & 3.036   & -1.08$\pm$0.06 &     89.3$\pm$4.9 &  20-4000  &  (2.69$\pm$0.54)$\times10^{-5} $& 20-4000 &K &1691$\pm$169  &  186$\pm$10  &   2259$\pm$453   &27 \\
080613B &	  & -1.05$\pm$0.18 &     22.1$\pm$4.1 &  20-3000  &  (4.76$\pm$1.31)$\times10^{-6}$ & 20-3000 &K & 733$\pm$239$^{\alph{footnote}}$\addtocounter{footnote}{-1} & & &28   \\
080721  & 2.591   & -0.96$\pm$0.07 [-2.42$\pm$0.29]&  88.1$\pm$7.6 &  20-7000  &  (2.11$\pm$0.35)$\times10^{-5}$ & 20-7000 &K &1785$\pm$223 & 121$\pm$10  &    1038$\pm$172 &29 \\
080804  & 2.20    & -0.72$\pm$0.04 & 	  9.88$\pm$0.35 &  8-35000 &  (7.30$\pm$0.88)$\times10^{-7}$ & 8-35000&F &  810$\pm$45  &  11.4$\pm$0.4 &    27.0$\pm$3.3	  &30 \\
080916A & 0.689   & -0.99$\pm$0.05 & 	  7.02$\pm$0.21 &  8-35000 &  (4.87$\pm$0.27)$\times10^{-7}$ & 8-35000 &F& 208$\pm$11   &  0.92$\pm$0.03 &    1.08$\pm$0.06	      &30 \\
081007  & 0.53    & -1.4$\pm$0.4   &      1.2$\pm$0.1  &  25-900   &  2.2$\pm$0.2$^{\alph{footnote}}$  & 25-900  &F &   61$\pm$15  &  0.17$\pm$0.015 &    0.43$\pm$0.04 &31 \\
081121  & 2.512   & -0.46$\pm$0.08 [-2.19$\pm$0.07]&  28.4$\pm$2.45&  8-35000 & (5.17$\pm$0.83)$\times10^{-6}$ & 8-35000&F & 608$\pm$42   &  30.5$\pm$2.6 & 195$\pm$31  &30 \\	    
081203A & 2.10    & -1.29$\pm$0.14 &    30.5$\pm$11.2&  20-3000  &  2.9$\pm$0.2$^{\alph{footnote}}$     & 15-150 &S/K& 1541$\pm$756 &  35.0$\pm$12.8 &    28.2$\pm$1.9   &32,33\\
081221  & 2.26  & -0.83$\pm$0.01  &  27.2$\pm$0.13 &  8-35000 &  (24.2$\pm$0.50)$\times10^{-7}$ & 8-35000&F &  284$\pm$2     &  34.6$\pm$0.2         &100$\pm$2 &30 \\
081222  & 2.77    & -0.90$\pm$0.03  [-2.33$\pm$0.10]&  17.6$\pm$1.63 &  8-35000 &  (17.6$\pm$0.58)$\times10^{-7}$ & 8-35000&F & 630$\pm$31 &  25.2$\pm$2.3  &  94.9$\pm$3.1 &30 \\
090102  & 1.547   & -0.97$\pm$0.01 &   34.9$\pm$0.66 &  8-35000 &  (29.3$\pm$0.91)$\times10^{-7}$ & 8-35000&F & 1174$\pm$38  &  21.4$\pm$0.4 &    45.7$\pm$1.4 &30 \\
090201  &$<$4     & -0.97$\pm$0.09 [-2.80$\pm$0.52]&  67.2$\pm$5.0&  20-2000  &  (7.30$\pm$1.26)$\times10^{-6}$ & 20-2000 &K & $<$790       &  $<$229       &    $<$1246 &34 \\
090424  & 0.544   & -1.02$\pm$0.01 [-3.26$\pm$0.18]&  50.1$\pm$1.06&  8-35000 &  (9.12$\pm$0.14)$\times10^{-6} $& 8-35000&F  & 250$\pm$3.4 & 3.97$\pm$0.08  & 11.2$\pm$0.17 &30 \\
\hline
\end{tabular}
\end{center}
\label{tab1}
\end{minipage}
\end{table*}

\begin{table*}
 \centering
 \begin{minipage}{176mm}
  \addtocounter{table}{-1}
 \caption{Continued}
 \addtocounter{footnote}{0}
 \footnotetext[\value{footnote}]{In these cases the peak flux is in units of ph/cm$^2$/s}
 \addtocounter{footnote}{1}
\footnotetext[\value{footnote}]{For GRBs without measured $z$ the listed peak energy refers to the observed one \epo.}
\addtocounter{footnote}{1}
\footnotetext[\value{footnote}]{The spectrum is well described by a power--law model and \epo\ is unconstrained.}
\addtocounter{footnote}{-2} 

\begin{center}
\tabcolsep 4.5pt
\begin{tabular}{@{}lcccccccccccl@{}}
  \hline
GRB & $z$ & $\alpha$ [$\beta$] & Fluence & Range & Peak Flux & Range &Mission& \ep\ & \eiso\ & \liso\  &Ref. \\
         &        &                &              10$^{-6}$erg/cm$^{2}$& keV  & erg/cm$^{2}$/s& keV & & keV&10$^{52}$erg&10$^{51}$erg/s& \\
  \hline
090709A & $<$3.5  & -0.85$\pm$0.08 [-2.7$\pm$0.24]&  91$\pm$0.7 &  20-3000  &  (3.9$\pm$0.6)$\times10^{-6}$  & 20-3000 &K  & $<$1341	  &  $<$229   &    $<$442   &35 \\
090715B & 3.00    & -1.1$\pm$0.37 &    9.3$\pm$1.3  &  20-2000  &  (9.0$\pm$2.5)$\times10^{-7}$  & 20-2000&K  & 536$\pm$164  &  21.3$\pm$3.0  &    82.6$\pm$22.9 &36 \\
090812  & 2.452   & -1.03$\pm$0.07    &  26.1$\pm$3.4&  15-1400  &  2.77$\pm$0.28$^{\alph{footnote}}$    & 100-1000 &Su/K/S &2023$\pm$663  &  40.5$\pm$5.3  &   96.2$\pm$9.7 &37,38\\
090926B & 1.24    & -0.19$\pm$0.06   &  9.81$\pm$0.16 &  8-35000 &  (4.73$\pm$0.28)$\times10^{-7}$ & 8-35000&F & 212$\pm$4.3  &  3.96$\pm$0.06  &    4.28$\pm$0.25  &30 \\
091018  & 0.971   & -1.53$\pm$0.48 &    1.44$\pm$0.17 &  20-1000  &  (4.32$\pm$0.95)$\times10^{-7}$ & 20-1000 &K  &  55$\pm$26   &  0.80$\pm$0.09  &    4.73$\pm$1.04 &39 \\
091020  & 1.71    & -1.20$\pm$0.06 [-2.29$\pm$0.18]&  12.4$\pm$1.81 & 8-35000 & (1.88$\pm$0.26)$\times10^{-6}$ & 8-35000 &F& 507$\pm$68   &  7.96$\pm$1.16  &    32.7$\pm$4.6 &30\\
091127  & 0.49    & -1.25$\pm$0.05 [-2.22$\pm$0.01]&  24.8$\pm$0.54 &  8-35000 &  (9.38$\pm$0.23)$\times10^{-6}$ & 8-35000 &F & 51$\pm$1.5 & 1.61$\pm$0.03  & 9.08$\pm$0.22 &30\\
091208B & 1.063   & -1.29$\pm$0.04 &  5.98$\pm$0.17 &  8-35000 &  (25.6$\pm$0.97)$\times10^{-7}$ & 8-35000&F & 246$\pm$15   &  1.97$\pm$0.06 &    17.4$\pm$0.7 &30 \\
100615A & 	  & -1.24$\pm$0.07 [-2.27$\pm$0.11]&  8.64$\pm$0.17 &  8-1000 &  8.3$\pm$0.2$^{\alph{footnote}}$\addtocounter{footnote}{1}     & 8-1000 &F& 86$\pm$8.5$^{\alph{footnote}}$\addtocounter{footnote}{1} &        & &40 \\
100621A & 0.542   & -1.70$\pm$0.13 [-2.45$\pm$0.15]&  36$\pm$4  &  20-2000  &  (1.70$\pm$0.13)$\times10^{-6}$ & 20-2000  &K& 146$\pm$23   &  4.35$\pm$0.48 & 3.17$\pm$0.24 &41 \\
100728B & 2.106   & -0.90$\pm$0.07    &  2.69$\pm$0.11 &   8-35000 &  (5.43$\pm$0.35)$\times10^{-7} $ & 8-35000&F  & 404$\pm$29   &  2.98$\pm$0.13  &    18.7$\pm$1.2 &42\\
110205A & 2.22    & -1.52$\pm$0.14   &  36.6$\pm$3.5 &  20-1200  &  (5.1$\pm$0.7)$\times10^{-7} $ & 20-1200 &K  & 715$\pm$238  &  55.9$\pm$5.3  &    25.1$\pm$3.4  &43 \\
110503A & 1.613   & -0.98$\pm$0.08 [-2.7$\pm$0.3] &  26$\pm$2 &  20-5000  &  (1.0$\pm$0.1)$\times10^{-5} $ & 20-5000&K  & 572$\pm$50   &  18.0$\pm$1.4 &    181$\pm$18  &44\\
\hline
\end{tabular}
\end{center}
\label{tab1}
\end{minipage}
\end{table*}

\begin{table*}
 \centering
 \begin{minipage}{176mm}
  \caption{
  Results of the statistical analysis on the \epeiso\ and \epliso\ correlations for (i) the total sample, (ii) the 
  complete sample presented in this work and (iii) the sample of bursts belonging to the total sample but not included 
  in the complete one (complementary sample). The table lists the Spearman's rank correlation coefficient ($\rho$), its 
  associated chance probability, the slope and normalization of the power--law fit and the 1$\sigma$ scatter of the 
  point distribution around the best fit line. 
  }
  \begin{center}
\begin{tabular}{llrclccc}
  \hline
  Correlation & Sample & \#GRBs & $\rho$  & P$_{\rm chance}$ &  Slope  &  Norm.  &  $\sigma_{\rm sc}$  \\
  \hline
                    & Total &  136 &0.77  &  4$\times10^{-28}$  &  0.55$\pm$0.02 & -26.74$\pm$1.13 & 0.23 \\
$E_{\rm peak}-E_{\rm iso}$ & Complete & 46& 0.76 & 7$\times10^{-10}$& 0.61$\pm$0.04 & -29.60$\pm$2.23 & 0.25\\
                    & Complementary& 90&0.78& 3$\times10^{-19}$ & 0.53$\pm$0.02& -25.55$\pm$1.35 & 0.25\\
   \hline
                    &Total & 135 &0.74 & 8$\times10^{-25}$ & 0.49$\pm 0.03$ & -22.98$\pm1.81$ & 0.30\\
$E_{\rm peak}-L_{\rm iso}$ & Complete & 46&0.65 & 1$\times10^{-6}$ & 0.53$\pm$0.06 & -25.33$\pm$3.26& 0.29\\
                    & Complementary & 89  &0.75 & 3$\times10^{-17}$ & 0.48$\pm$0.04 & -22.44$\pm$2.12 &0.30\\
 \hline
\hline
\end{tabular}
\end{center}
\label{tab2}
\end{minipage}
\end{table*}

\subsection{Distributions of \ep, \eiso\ and \liso}
The complete sample adopted in this work has been well selected and the effects induced by the limiting peak flux have been well understood and quantified in Sect. 3.1. On the other hand, up to now, the spectral--energy correlations have been studied by means of a sample which includes all possible GRBs with known redshift and \ep, detected by different instruments. This total sample, therefore, is affected by several instrumental effects that can play important roles (as the different detector sensitivities and the different energy ranges on which the detection is performed). To estimate these effects and to understand how they influence the properties of the total sample is not trivial. 
Here we can use a different approach and point out possible biases by comparing the results obtained from the total sample with those from the complete one. From the point of view of correlations, we demonstrated that the total sample does not introduce evident biases on their slopes, normalizations and scatters (see Table \ref{tab2}). However, this does not necessarily imply that the complete and total samples share the same properties in terms of distributions of \ep, \eiso\ and \liso.

\begin{figure}
\hskip 0.3 truecm
\includegraphics[scale=0.55]{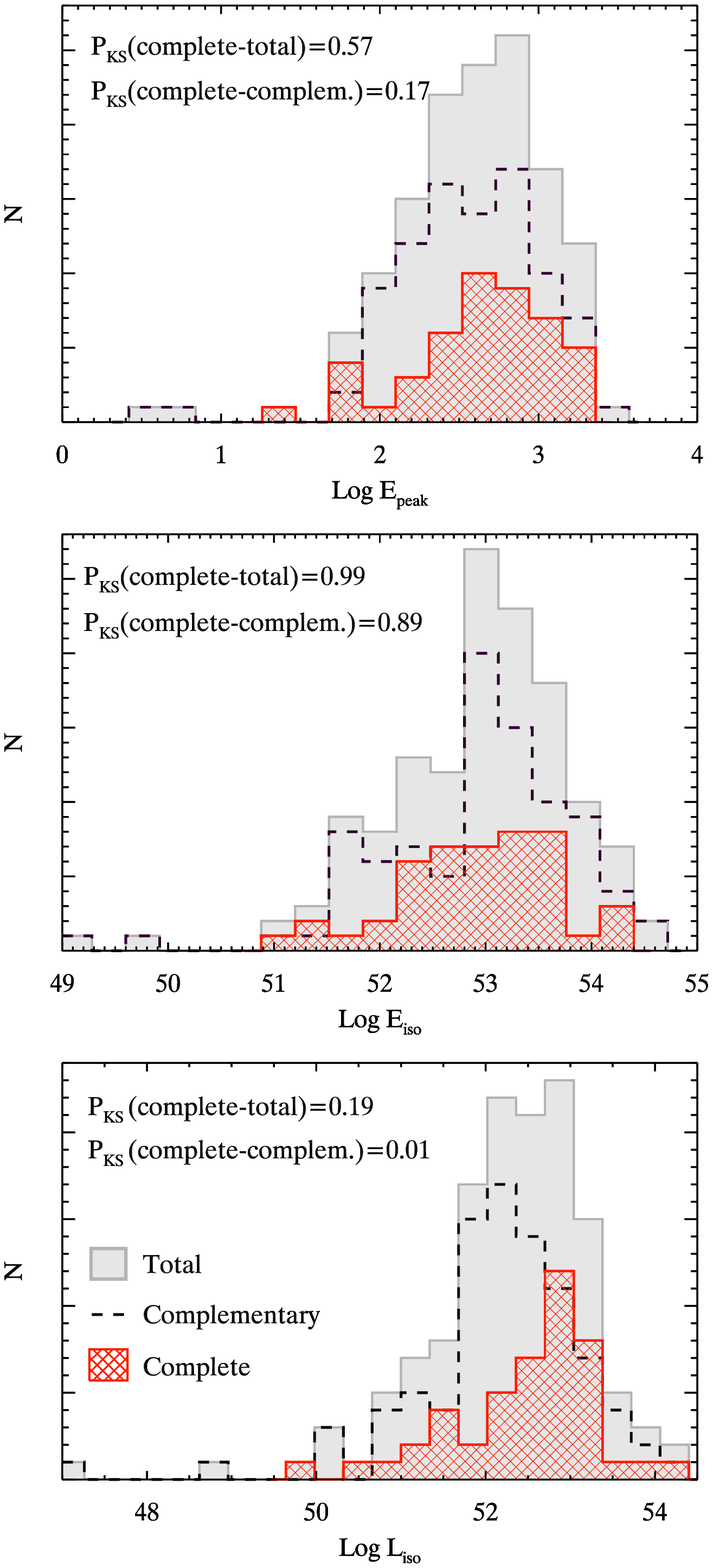} 
 \caption{Distributions of \ep\ (rest frame), \eiso\ and \liso\ for bursts of the total (filled grey histograms), complementary (dashed dark line) and complete (hatched red histograms) sample (see legend in the bottom panel). In each panel the probability of the K-S test is also reported.}
 \label{fig4}
\end{figure}

We investigate possible effects present in the total sample by comparing the \ep, \eiso\ and \liso\ distributions of the complete, total and complementary sample.
For each quantity, we also apply the Kolmogorov-Smirnov (K-S) test and estimate the probability that two 
distributions are drawn from the same parent population. 
In particular, we apply this test to compare i) the 
complete and total sample and ii) the complete and complementary sample. 
Results are shown in Fig. \ref{fig4}. 
In all cases, the K-S probability between the complete and the total sample is large, and the distributions are 
consistent within 1$\sigma$. 
The probability slightly decreases when comparing the complete and the complementary sample. 
In this case all distributions are consistent within 2$\sigma$. 
A slight effect is possibly present in the \liso\ distribution, where the selection cut discussed in the previous 
section lowers the number of intermediate/low luminosity events. 
Due to the relation between \liso\ and \ep, this effect also influence the \ep\ distribution. Even if numbers are still small for the complete sample, our results suggest that the total sample is not affected by any relevant bias.

\section{Discussion and conclusions}

The definition of a sample of GRBs with a high level of completeness (90\%) allows to 
address several open issues regarding population studies of GRBs. 
In this paper we use this complete sample to study the spectral--energy correlations in an unbiased way. 
Out of 58 GRBs satisfying the selection criteria 46 have well 
defined spectral properties and measured redshift. 
In particular, the rest frame peak energy \ep\ can be determined and its relation with \eiso\ and \liso\ 
can be investigated. 
We find that this sample defines two strong 
correlations: (i) \epeiso\ (rank's correlation coefficient $\rho=0.76$ and chance probability 
P$_{\rm chance}=7\times 10^{-10}$) and (ii) \epliso\ ($\rho=0.65$ and P$_{\rm chance}=1\times 10^{-6}$).

The slope of the correlations defined by the complete sample are 0.61 and 0.53 for the Amati and the Yonetoku 
correlations, respectively. 
These slopes are consistent within 1$\sigma$ (2$\sigma$) with the slopes of the \epliso\ (\epeiso) 
correlation defined by bursts not satisfying the completeness criteria (complementary sample). 
From Table \ref{tab2} we also note that the scatter of the correlations defined with the complete, total and 
complementary sample are similar among themselves and are about $\sigma_{\rm sc}=0.25$ and 
$\sigma_{\rm sc}=0.30$ for the Amati and Yonetoku correlation. 
We also investigated possible instrumental effects in the correlations found in this complete sample. We defined three sub-samples based on the instrument which derived the burst spectral parameters (see Table 1). In particular, we study the Amati and Yonetoku correlations within the sample of Swift/BAT bursts (10 events), Konus bursts (18 events) and Fermi/GBM bursts (13 events). 
The results derived from different sub-samples are very similar and consistent within the errors.
In fact, we found that the best fit slopes (normalizations) of the Amati correlation are 0.60$\pm$0.05 
(-28.97$\pm$2.67),  0.59$\pm$0.08 (-28.41$\pm$4.19) and 0.60$\pm$0.11 (-29.19$\pm$5.77) for the Swift, Konus and Fermi samples respectively. For the Yonetoku 
correlation we find: 0.44$\pm$0.11 (-20.76$\pm$5.95), 0.52$\pm$0.08 (-24.61$\pm$4.49) and 0.54$\pm$0.12 (-25.87$\pm$6.37).

Our study outlines the presence of one GRB (061021) that lies at 3$\sigma$ limit (or more) of the tested correlations. 
In particular, it is an outlier to the Amati correlation.

GRB 061021 has been detected by {\it Swift}/BAT (Moretti et al. 2006) and by {\it Konus}/WIND 
(Golenetskii et al. 2006). It shows a single pulse with a duration of $\sim10$ s, followed by a weak tail seen 
up to $\sim60$ s. 
On the basis of the lag analysis (Norris \& Barthelmy 2006) this burst is classified as a long duration event.
The time integrated BAT spectrum is well described by a power--law model with photon index 
--1.31$\pm 0.06$ (Palmer et al. 2006), suggesting \epo$>150$ keV. 
Golenetskii et al. (2006) report the spectral 
analysis of {\it Konus}/WIND data separately for the pulse and the tail. 
The pulse is well described by a power--law with a high--energy exponential cutoff, 
with spectral parameters $\alpha=-1.22^{+0.12}_{-0.14}$ and \epo
$=777^{+549}_{-237}$. The soft tail can be modelled as a single power--law, 
with photon index $-1.93^{+0.32}_{-0.27}$ and a fluence which is nearly 1/3 of the fluence of the main pulse. 
In order to avoid to underestimate the total energetics, we also include the fluence of the soft tail in the 
computation of \eiso. Nevertheless, this burst is on average $\sim$2 orders of magnitude less energetic as compared to 
bursts characterized by a similar peak energy. Note that the lack of a spectral analysis integrated over the whole burst duration does not affect the estimate of \liso, which is well computed by using the peak flux and the spectral shape of the main peak.  Also in this case, the peak luminosity is $\sim$2 orders of magnitude less than the average \liso\ of events with \ep\ similar to GRB 061021.
Finally, the redshift of this source is $z=0.346$ (Fynbo et al. 2009).

The presence of outliers to the Amati relation is somehow expected from the study of the observational 
plane \epo$-Fluence$. Nava et al. (2008; 2011b) studied the distribution of BATSE and GBM bursts in this 
plane and derived respectively a 6\% and 3\% of outliers. 
We also stress that the definition of the outliers of the \epeiso\ correlation refers to the present knowledge of the 
scatter of this correlation which is assumed to be Gaussian. It might well be that the scatter is still not fully known 
and/or that it is not Gaussian. The latter possibility seems to be suggested also by the log--normal distribution of the jet opening angles
(Ghirlanda et al. 2005) if the source of the scatter of the \epeiso\ correlation were, indeed, the jet opening angle.

There are 12 bursts belonging to the complete sample that cannot be used to directly test 
the correlations, due to the lack of measured redshift and/or \epo. 
However, we can still check their consistency with the correlations (Fig. \ref{fig2}). 
We find that they are all consistent with both correlations. 

By taking advantage of our complete sample, for both correlations we investigate the possible evolution with the redshift of the best fit slopes (Fig. \ref{fig3}). To this aim we define 4 bins of redshift, chosen in order to have 4 sub--samples characterized by a similar number of objects. We find no relation between the slope of each sub--sample and the redshift. All slopes are consistent with the one defined by the whole complete sample and this is true both for the \epeiso\ and the \epliso\ correlation.

We also investigate possible differences (in terms of distributions of the most fundamental quantities) 
between bursts belonging to the complete sample and bursts that do not satisfy the complete sample selection criteria. Our aim is to understand if the different instrumental biases present in the total sample affect in some way the distributions of the prompt emission properties.
We find that for all the considered quantities (\ep, \eiso\ and \liso) the distributions derived from the complete and from the complementary samples are consistent within 2$\sigma$.

Our results give support to the idea that spectral energy correlations arise because
of a robust physical mechanism common to the vast majority of bursts.
Outliers exist, but they are a few. Furthermore, the evidence
that GRBs do evolve in redshift (Salvaterra \& Chincarini 2007), while there is no
sign of evolution of the slopes of the correlations, also suggests that 
the spectral peak energy is
closely linked to the luminosity/energetics of the burst, another sign that selection effects are
not important.

\section*{Acknowledgments}
We thank the referee for comments that improved the manuscript.
This work has been supported by ASI grant I/004/11/0 and ASI grant I/088/06/0.
This research has made use of the public Fermi/GBM data 
and software obtained through the High Energy Astrophysics Science Archive 
Research Center Online Service, provided by the NASA/Goddard Space Flight 
Center.

\label{lastpage}

\end{document}